\documentclass[aps,prl,twocolumn,superscriptaddress,reprint]{revtex4-2}
\usepackage{graphicx}
\usepackage{amssymb, amsmath}
\usepackage{booktabs}
\usepackage[usenames]{color}
\usepackage{bm}
\usepackage{multirow}
\usepackage{dcolumn}
\usepackage{hyperref}
\usepackage{enumerate}
\hypersetup{
    colorlinks=true,
    linkcolor=blue,
    filecolor=gray,      
    urlcolor=blue,
    citecolor=blue,
}

\begin{document}

\title{Deep-learning electronic-structure calculation of magnetic superstructures}

\affiliation{State Key Laboratory of Low Dimensional Quantum Physics and Department of Physics, Tsinghua University, Beijing, 100084, China}
\affiliation{Tencent Quantum Laboratory, Tencent, Shenzhen, Guangdong 518057, China}
\affiliation{Institute for Advanced Study, Tsinghua University, Beijing 100084, China}
\affiliation{School of Physics, Peking University, Beijing 100871, China}
\affiliation{Frontier Science Center for Quantum Information, Beijing, China}
\affiliation{RIKEN Center for Emergent Matter Science (CEMS), Wako, Saitama 351-0198, Japan}

\author{He \surname{Li}}
\affiliation{State Key Laboratory of Low Dimensional Quantum Physics and Department of Physics, Tsinghua University, Beijing, 100084, China}
\affiliation{Tencent Quantum Laboratory, Tencent, Shenzhen, Guangdong 518057, China}
\affiliation{Institute for Advanced Study, Tsinghua University, Beijing 100084, China}

\author{Zechen \surname{Tang}}
\affiliation{State Key Laboratory of Low Dimensional Quantum Physics and Department of Physics, Tsinghua University, Beijing, 100084, China}

\author{Xiaoxun \surname{Gong}}
\affiliation{State Key Laboratory of Low Dimensional Quantum Physics and Department of Physics, Tsinghua University, Beijing, 100084, China}
\affiliation{School of Physics, Peking University, Beijing 100871, China}

\author{Nianlong \surname{Zou}}
\affiliation{State Key Laboratory of Low Dimensional Quantum Physics and Department of Physics, Tsinghua University, Beijing, 100084, China}

\author{Wenhui \surname{Duan}}
\email{duanw@tsinghua.edu.cn}
\affiliation{State Key Laboratory of Low Dimensional Quantum Physics and Department of Physics, Tsinghua University, Beijing, 100084, China}
\affiliation{Tencent Quantum Laboratory, Tencent, Shenzhen, Guangdong 518057, China}
\affiliation{Institute for Advanced Study, Tsinghua University, Beijing 100084, China}
\affiliation{Frontier Science Center for Quantum Information, Beijing, China}

\author{Yong \surname{Xu}}
\email{yongxu@mail.tsinghua.edu.cn}
\affiliation{State Key Laboratory of Low Dimensional Quantum Physics and Department of Physics, Tsinghua University, Beijing, 100084, China}
\affiliation{Tencent Quantum Laboratory, Tencent, Shenzhen, Guangdong 518057, China}
\affiliation{Frontier Science Center for Quantum Information, Beijing, China}
\affiliation{RIKEN Center for Emergent Matter Science (CEMS), Wako, Saitama 351-0198, Japan}

\begin{abstract}
{\it Ab initio} study of magnetic superstructures (e.g., magnetic skyrmion) is indispensable to the research of novel materials but bottlenecked by its formidable computational cost. For solving the bottleneck problem, we develop a deep equivariant neural network method (named xDeepH) to represent density functional theory Hamiltonian $H_\text{DFT}$ as a function of atomic and magnetic structures and apply neural networks for efficient electronic structure calculation. Intelligence of neural networks is optimized by incorporating {\it a priori} knowledge about the important locality and symmetry properties into the method. Particularly, we design a neural-network architecture fully preserving all equivalent requirements on $H_\text{DFT}$ by the Euclidean and time-reversal symmetries ($E(3) \times \{I,\mathcal{T}\}$), which is essential to improve method performance. High accuracy (sub-meV error) and good transferability of xDeepH are shown by systematic experiments on nanotube, spin-spiral, and Moir\'{e} magnets, and the capability of studying magnetic skyrmion is also demonstrated. The method could find promising applications in magnetic materials research and inspire development of deep-learning {\it ab initio} methods. 
\end{abstract}
\maketitle

The subject of magnetic superstructures, such as magnetic skyrmion, Moir\'{e} magnetism, and spin-spiral magnet, has attracted intensive research interest, which opens opportunities to explore emergent physics in quantum materials, including skyrmion Hall effect, topological Hall effect, flat-band physics, etc~\cite{Fert2013, Nagaosa2013}. {\it Ab initio} calculation based on density functional theory (DFT) has become an indispensable tool for the research, but is only applicable to the study of small-scale superstructures due to the demanding computational cost. Recent developments of deep learning {\it ab initio} methods~\cite{Carleo2019, Behler2007, Zhang2018, Justin2017, Schutt2018, Xie2018, Unke2021_2, Schutt2019, Li2022,Li2022_2, Gong2022, Unke2021,Su2022,Zhong2022, Batzner2022, Musaelian2022, Klicpera2020} shed light on solving the bottleneck problem, which use artificial neural networks to learn from {\it ab initio} data and apply neural networks for material simulation without invoking {\it ab initio} codes, enabling the study of large-scale material systems. However, current methods are usually designed to treat systems without magnetism, which neglect the dependence of material property on magnetic structure, thus not suitable for the purpose.

\begin{figure}
\includegraphics[width=1\linewidth]{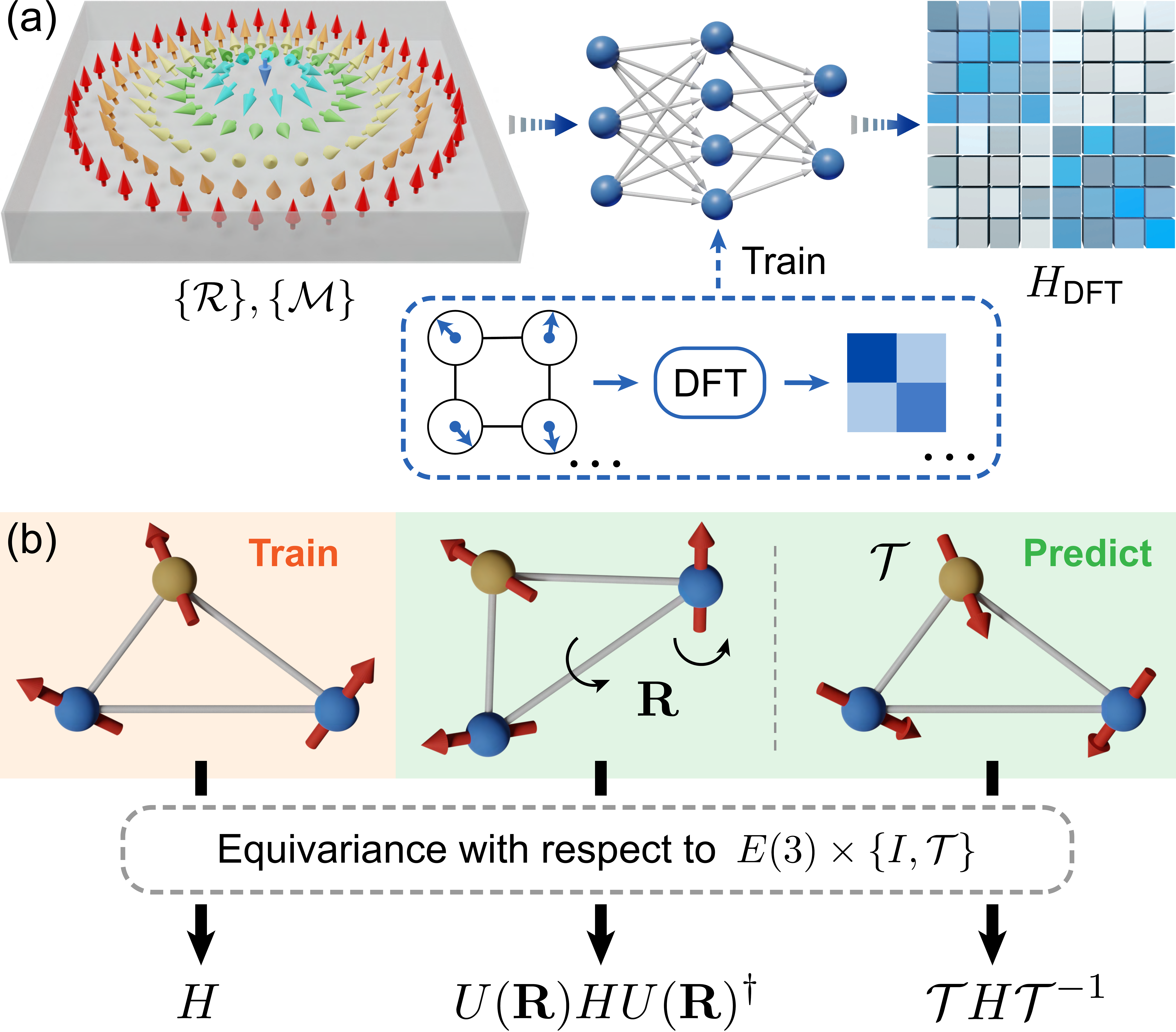}
\caption{Extended deep-learning DFT Hamiltonian (xDeepH) method for studying magnetic materials. (a) Workflow of the xDeepH method. Deep neural networks are used to represent DFT Hamiltonian $H_{\text{DFT}}$ as a function of atomic structure $\{\mathcal R\}$ and magnetic structure $\{\mathcal M\}$. The neural-network models are trained by DFT data on small-size structures and then applied to study magnetic superstructures, such as magnetic skyrmion. (b) Equivalence of $H_\text{DFT}\left(\{\mathcal R\}, \{\mathcal M\}\right)$ with respect to the $E(3)\times\{I,\mathcal T\}$ group. Transformations of rotation $\mathbf R$ and time-reversal $\mathcal T$ are illustrated. Colored balls and arrows denote atoms and magnetic moments, respectively.}
\label{fig1}
\end{figure}

A key problem of deep learning DFT calculation is to design deep neural network models to represent the DFT Hamiltonian $H_\text{DFT}$ as a function of atomic structure $H_\text{DFT}\left(\{\mathcal R\}\right)$ for efficient electronic-structure simulation~\cite{Li2022,Li2022_2}. The problem has recently been investigated for nonmagnetic systems~\cite{Li2022,Li2022_2,Gong2022,Unke2021,Schutt2019,Su2022,Zhong2022}. The counterpart problem for magnetic systems is of equal importance, which looks more complicated on the following aspects: First, an extra dependence on magnetic structure $\{\mathcal M\}$ is introduced into $H_\text{DFT}$ [Fig. \ref{fig1}(a)], which is physically distinct from the dependence on $\{\mathcal R\}$. Second, the spin degrees of freedom are usually negligible in the nonmagnetic case, but become essential here.  Consequently $H_\text{DFT}$ gets non-diagonal in spin space, leading to an enlarged number of nonzero matrix elements [Fig. \ref{fig1}(a)]. Third, satisfying fundamental symmetry conditions is a prerequisite for achieving good performance in the deep learning problem~\cite{Li2022,Li2022_2,Gong2022,Unke2021}. Generalized symmetry requirements on neural network models of $H_\text{DFT}$ are imposed by symmetry operations on both $\{\mathcal M\}$ and $\{\mathcal R\}$. This critical issue has not been addressed before. Actually, the nonmagnetic problem is a limiting case of the magnetic problem with vanishing $\{\mathcal M\}$. In this context, significant generalization of deep learning DFT methods is possible and urgently demanded by the research field.

In this work, we develop an extended deep-learning DFT Hamiltonian (xDeepH) method, including theoretical framework, numerical algorithm and computational code, which learns the dependence of spin-polarized DFT Hamiltonian on atomic and magnetic structures by deep equivariant neural network (ENN) models, enabling efficient electronic structure calculation of large-scale magnetic materials. As a critical innovation, we design an ENN architecture to incorporate physical insights and respect the fundamental symmetry group $E(3) \times \{I,\mathcal{T}\}$ (Euclidean and time-reversal symmetries) of $H_\text{DFT}\left(\{\mathcal R\}, \{\mathcal M\}\right)$, ensuring efficient and accurate deep learning. The method is systematically tested to show high precision (sub-meV error) and good transferability by example studies of magnetic superstructures ranging from nanotube magnets, spin-spiral magnets, to magnetic skyrmions. Benefiting from the extended capability and state-of-the-art performance, xDeepH could find promising applications in materials research and stimulate development of deep learning {\it ab initio} methods.

Deep-learning DFT Hamiltonian (DeepH) method has been developed to improve efficiency of electronic structure calculation, which shows great potential to address the accuracy-efficiency dilemma of DFT~\cite{Li2022,Li2022_2}. A significant generalization of the method is required to study a broad class of magnetic materials. For nonmagnetic systems, $H_\text{DFT}$ as a function atomic structure $\{\mathcal R\}$ is calculated by self-consistent field (SCF) iterations in DFT~\cite{Martin2004}. The function $H_\text{DFT}\left(\{\mathcal R\}\right)$ is the learning target of DeepH. In contrast, for magnetic systems $H_\text{DFT}$  depends not only on atomic structure but also on magnetic structure $\{\mathcal M\}$. To compute $H_\text{DFT}$ for a given $\{\mathcal M\}$, one needs to apply constrained DFT that employs Lagrangian approach to constrain magnetic configuration and introduces constraining fields into the Kohn-Sham potential~\cite{Dederichs1984}. Generally the mapping from $\{\mathcal R\}$ and $\{\mathcal M\}$ to the spin-polarized $H_\text{DFT}$ is unique in constrained DFT~\cite{Wu2005}. $H_\text{DFT}$ is also calculated by SCF iterations but demands much more computational resource than the nonmagnetic counterpart. This is because the additional constraining fields should be determined self-consistently, and an enlarged Hamiltonian matrix non-diagonal in spin space must be considered.

The workflow of xDeepH is illustrated in Fig. \ref{fig1}(a). First, small-size materials with diverse atomic and magnetic configurations are calculated by constrained DFT for preparing datasets. Then deep neural networks representing $H_\text{DFT}\left(\{\mathcal R\}, \{\mathcal M\}\right)$ are trained on the datasets. Next, the neural networks are applied to predict $H_\text{DFT}$ for materials with varying atomic and magnetic structures. Based on $H_\text{DFT}$, any electronic properties of materials in the single-particle picture can be computed. By replacing DFT SCF calculation with deep neural networks, xDeepH greatly reduces the computational cost of electronic structure calculation and enables the study of magnetic superstructures (e.g., magnetic skyrmions). The critical issue, however, is the design of intelligent neural networks for modelling the mapping function $\left(\{\mathcal R\}, \{\mathcal M\}\right)\mapsto H_{\text{DFT}}$, using as much {\it a priori} knowledge as possible for optimizing neural-network performance.

Two physical principles are essential to the deep learning problem here, including the nearsightedness (or locality) principle of electronic matter and the symmetry principle. Local physical properties satisfying the nearsightedness principle are insensitive to distant change of chemical environment~\cite{Prodan2005}. For instance, charge density belongs to local property, whereas the Kohn-Sham eigenstates are nonlocal. The latter depends sensitively on the global material structure, which is complicated from the point view of machine learning. In general, qualities with local property are more favorable for deep learning than nonlocal ones. Moreover, the fundamental physical laws are covariant under symmetry operations (e.g., translation, rotation). The symmetry is an inherent property of physical data. Thus the use of symmetry properties could significantly facilitate deep learning. In short, the principles of locality and symmetry are {\it a priori} knowledge of pivotal importance to artificial intelligence.

Let us first check the locality nature of the deep learning target $H_\text{DFT}\left(\{\mathcal R\}, \{\mathcal M\}\right)$. In DFT calculations, plane waves and localized orbitals are usually employed as basis functions. The latter kind of basis is preferred, as it is compatible with the locality principle. The orbital functions have the form $\phi_{iplm}(\boldsymbol{r})=R_{i p l}(r) Y_{l m}(\hat{r})$, where the radial function $R_{i p l}(r)$ is centered at the $i_{\rm{th}}$ atom, $p$ is the multiplicity index, and the angular part $Y_{lm}(\hat{r})$ is the spherical harmonics of degree $l$ and order $m$. Herein the spin degree of freedom must be considered, which are labeled by $\sigma = \pm 1/2$. The matrix element is then written as $ [H_{ij}]^{p_1p_2;l_1 l_2}_{m_1\sigma_1, m_2\sigma_2}$, where the subscript of $H_\text{DFT}$ is omitted for simplicity. $H_{ij}$ is the Hamiltonian matrix block describing hopping between atoms $i$ and $j$. A notation of Hamiltonian matrix sub-block $\mathbf h \equiv [H_{ij}]^{p_1p_2}$ is introduced, whose elements have the form $\mathbf h^{l_1 l_2}_{m_1\sigma_1, m_2\sigma_2}$. Under the localized basis, the DFT Hamiltonian can be viewed as an {\it ab initio} tight-binding Hamiltonian. $H_{ij}$ has vanishing values for atom pairs with distance $d_{ij} > R_C$. The cutoff radius $R_C$ is determined by the spread of orbital functions and usually on the order of a few Angstroms. Importantly, $H_{ij}\left(\{\mathcal R\}, \{\mathcal M\}\right)$ obeys the nearsightedness principle, which is only influenced by changes of chemical environment of finite range $R_N$.

Noticeably, two kinds of nearsightedness length scales ($R_{N1}$ and $R_{N2}$) are relevant to influence induced by changes of $\{\mathcal R\}$ and $\{\mathcal M\}$, respectively. Formally, varying $\{\mathcal R\}$ will alter the strong external potential in $H_\text{DFT}$, whereas varying $\{\mathcal M\}$ will mainly modify the relatively weak constraining fields, leading to minor influence on $H_\text{DFT}$. It is thus expected that the latter influence on $H_\text{DFT}$ is much weaker in magnitude and shorter in length scale ($R_{N2} < R_{N1}$). This is confirmed by our numerical experiments (see Supplementary Material (SM)~\cite{supp}). Our results suggest that $R_{N2} \approx R_{C}$, and $R_{N1}$ is typically several times larger than $R_C$. The two distinct dependence behaviors of $H_\text{DFT}\left(\{\mathcal R\}, \{\mathcal M\}\right)$ should be accurately described together by the deep learning method. This important issue will be addressed in the dataset preparation and neural-network design.

Regarding the importance of symmetry in physics and remarkable performance of ENNs in prediction of materials properties~\cite{Li2022,Gong2022,Unke2021,Batzner2022,Musaelian2022,Klicpera2020}, it is essential to respect equivariance conditions in neural networks. The mapping function $H_\text{DFT}\left(\{\mathcal R\}, \{\mathcal M\}\right)$ considered in this work is equivariant with respect to the $E(3) \times \{I,\mathcal{T}\}$ group. If an overall rotation $\mathbf R$ is applied to atomic and magnetic structures, the Hamiltonian matrix sub-blocks will transform accordingly:
\begin{align}
\label{H_rot}
&\mathbf h^{l_1 l_2}_{m_1\sigma_1, m_2\sigma_2}\stackrel{\mathbf R}{\longrightarrow}\\
&\sum_{m_1',\sigma_1', m_2',\sigma_2'}D^{l_1\otimes\frac{1}{2}}_{m_1\sigma_1, m_1'\sigma_1'}(\mathbf R) \mathbf h^{l_1 l_2}_{m_1'\sigma_1', m_2'\sigma_2'}
\left[D^{l_2\otimes\frac{1}{2}}_{m_2\sigma_2,m_2'\sigma_2'}(\mathbf R)\right]^*,\nonumber
\end{align}
where the superscript ``$*$'' denotes the complex conjugate, and $D^{l}(\mathbf R)$ is the Wigner D-matrix.

\begin{figure}
\includegraphics[width=1\linewidth]{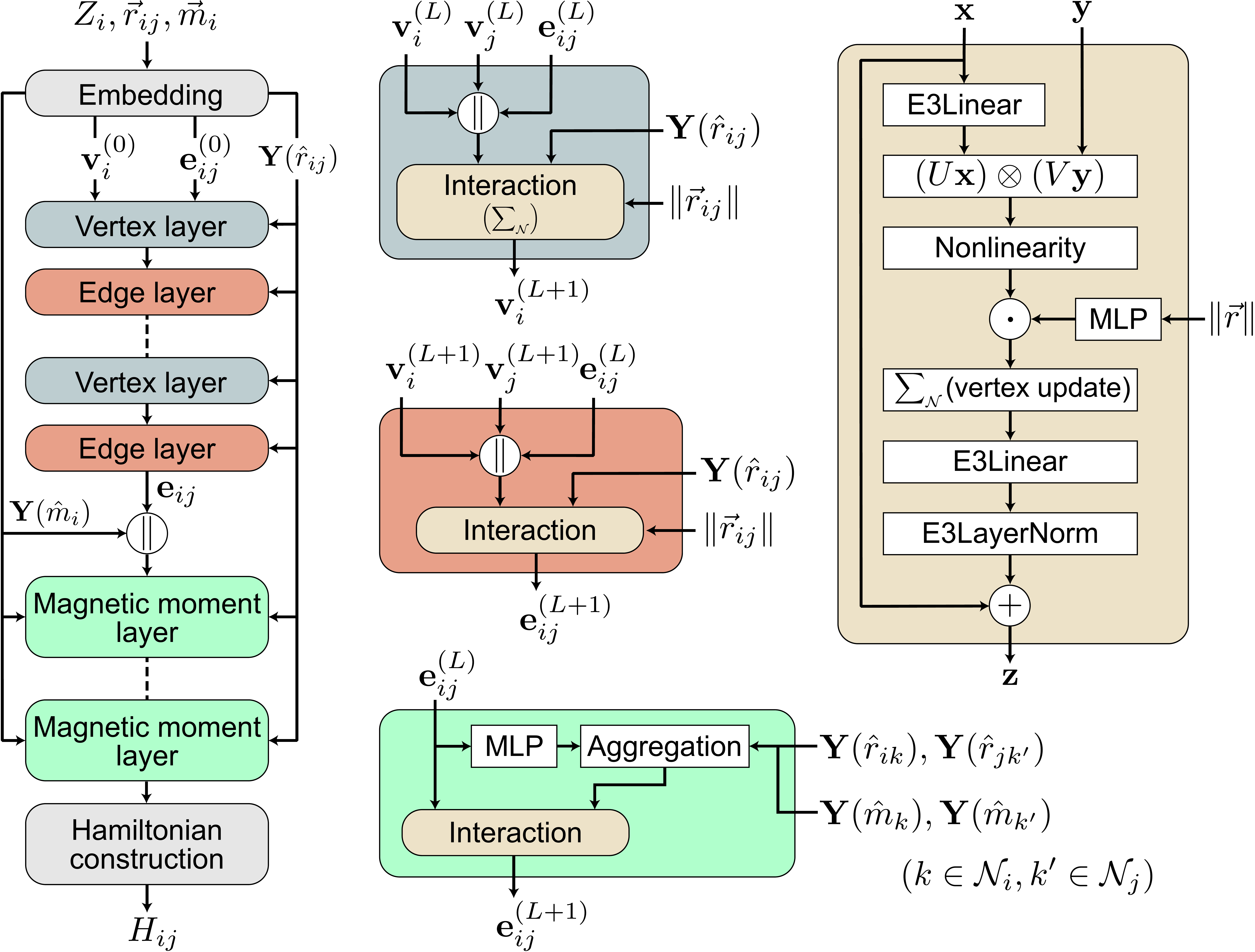}
\caption{Neural-network architecture of xDeepH. Initial vertex and edge features are embedded by one-hot encoding and Gaussian expansion, respectively. Features are updated alternately by vertex layer and edge layer with interatomic distance vectors $\vec r_{ij}$ equipped with spherical harmonics $\mathbf Y_{lm}$. ``$\|$'' denotes vector concatenation and ``$\cdot$'' denotes element-wise multiplication. ``$\sum_{\mathcal N}$'' denotes the summation of neighbors for features, which is only valid for the vertex layer. Subsequently, a magnetic moment layer is used to introduce the magnetic moments $\vec m_i$ of atoms as input in a strictly localized manner. More details are described in SM~\cite{supp}.}
\label{fig2}
\end{figure}

Equivariance with respect to the time-reversal operation $\mathcal T$ needs a special treatment. Under $\mathcal T$, the local magnetic moment of each atom is reversed, and $H_{\text{DFT}}$ is transformed by an antiunitary operator [Fig. \ref{fig1}(b)]. Since localized atomic orbitals are real functions, the time-reversal operation works only on the spin degrees of freedom of Hamiltonian matrix sub-blocks:
\begin{align}
\label{tr}
\mathbf h^{l_1 l_2}_{m_1\sigma_1, m_2\sigma_2}\stackrel{\mathcal T}{\longrightarrow}
\left\{
\begin{array}{lll}
&\left(\mathbf h^{l_1 l_2}_{m_1(-\sigma_1), m_2(-\sigma_2)}\right)^* \  \sigma_1 = \sigma_2\\
-&\left(\mathbf h^{l_1 l_2}_{m_1(-\sigma_1), m_2(-\sigma_2)}\right)^* \  \sigma_1 \neq \sigma_2.
\end{array}
\right.
\end{align}

The rotation equivariance can be achieved in the framework of ENNs~\cite{Thomas2018,Geiger2022}. ENNs learn the equivariant feature: if a rotation $\mathbf R$ is applied to the input coordinates of ENNs, the equivariant feature $\mathbf x_m^{l}$ will transform accordingly: $\mathbf x^{l}_{m}\stackrel{\mathbf R}{\longrightarrow}\sum_{m'} D^{l}_{mm'}(\mathbf R)\mathbf x^{l}_{m'}$, where $l$ marks that $\mathbf x_m^{l}$ carries the irreducible representation of the SO(3) group of dimension 2$l$+1, and $m$ is an integer between $-l$ and $l$. Feature vectors can be used to construct an equivariant matrix via $\mathbf X^{l_1l_2}_{m_1m_2}=\sum_{l_3,m_3}C_{m_1m_2m_3}^{l_1l_2l_3}\mathbf x^{l_3}_{m_3}$, where $C_{m_1m_2m_3}^{l_1l_2l_3}$ are Clebsch-Gordan coefficients. The feature matrices can be used to represent the Hamiltonian matrix sub-blocks in Eq. (\ref{H_rot}) because they follow the same transformation rule when the spin degrees of freedom are not included.

The time-reversal equivariant relation in Eq. (\ref{tr}) seems difficult to handle in the original ENN framework: matrix elements from different spin components are swapped and negative signs appear depending on diagonal or off-diagonal sub-blocks and real or imaginary parts. Moreover, due to the consideration of non-collinear spins, complex-valued equivariant features carrying irreducible representation of half-integer $l$ need to be constructed by neural networks. DeepH-E3 approch~\cite{Gong2022} performs a basis transformation for the output features of ENNs with only integer $l$ into the desired spin-$1/2$ equivariant features to deal with the spin-orbit coupling. We find that the same transformation can help solve the above issues. We demonstrate that all the desired output features of ENNs become time-reversal-even or -odd with the above basis transformation (see details in SM~\cite{supp}). These time-reversal-odd or -even equivariant features can be used to construct Hamiltonian sub-blocks that preserve Eq. (\ref{tr}). To extend ENNs to deal with this additional equivariance, we introduce an index $t$ into equivariant vectors $\mathbf x^{l;t}_{m}$ to mark its parity with respect to time-reversal. In addition to the rotational equivariance, the extended equivariant vectors should show the parity upon time-reversal: $\mathbf x^{l,t}_{m}\stackrel{\mathcal T}{\longrightarrow}(-1)^{t}\mathbf x^{l,t}_{m'}$,
where $t$ is 0 or 1 for time-reversal even or odd. The equivariance of spatial inversion is also handled in a similar and simpler way, since the DFT Hamiltonian matrix sub-block $\mathbf h^{l_1 l_2}$ natively has parity about spatial inversion, which is $(-1)^{l_1 + l_2}$. An additional parity index needs to be introduced into the equivariant vector to label the spatial-inversion even or odd.

\begin{figure*}
\includegraphics[width=1\linewidth]{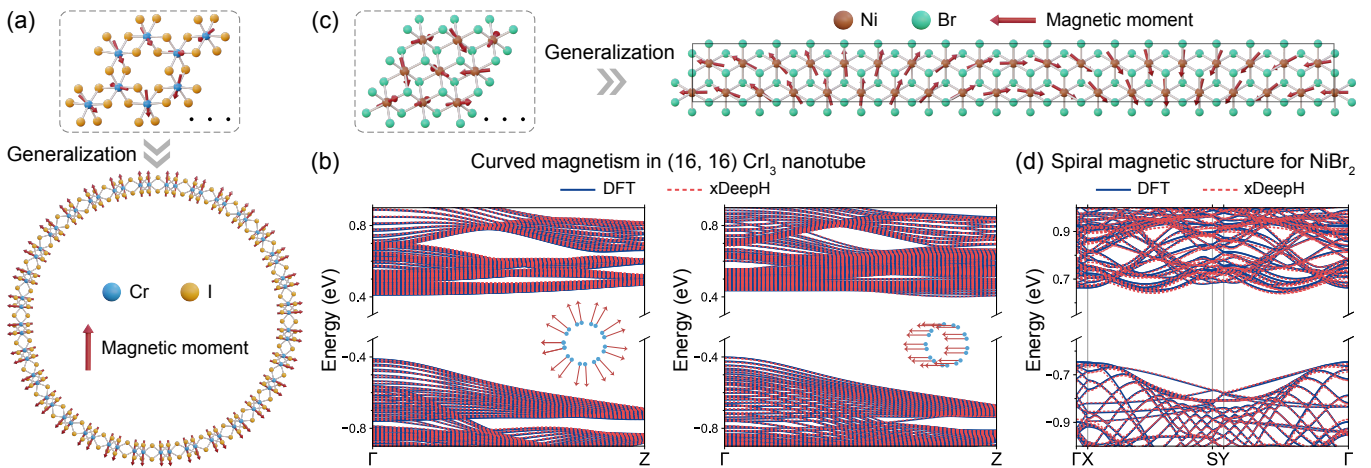}
\caption{Application of xDeepH to study nanotube and spin-spiral magnets. (a) Schematic workflow of xDeepH in the study of CrI$_3$, which uses DFT data on monolayer structures (32 atoms per supercell) for training and then make predictions on nanotubes. (b) Band structures of (16, 16)  CrI$_3$ nanotubes with the non-collinear magnetization normal to the surface (left) or collinear ferromagnetism (right) computed by DFT and xDeepH. (c) Schematic workflow of xDeepH in the study of NiBr$_2$, which uses DFT data on monolayer structures (27 atoms per supercell) for training and then make predictions on spiral magnetic structures. (d) Band structures of monolayer NiBr$_2$ with 19$\times$1$\times$1 spiral magnetic structure computed by DFT and xDeepH.}
\label{fig3}
\end{figure*}

Here we describe the ENN architecture for xDeepH, which fully preserves physical symmetries of the DFT Hamiltonian for magnetic materials. xDeepH is based on message-passing neural network (MPNN)~\cite{Gilmer2017}. The input atomic and magnetic structures are represented by a graph. Each vertex in the graph corresponds to an atom $i$ associated with its nuclear charge $Z_i$ and magnetic moment $\vec m_i$. Edges correspond to atom pairs $ij$ with nonzero $H_{ij}$. The interatomic distance vector $\vec r_{ij}$ for each edge is also taken as the input. $\vec r_{ij}$ are marked even under time-reversal and odd under spatial inversion, whereas $\vec m_i$ are marked odd under time-reversal and even under spatial inversion. The scalar input $Z_i$ is used to construct the equivariant feature with $l=0$. Real spherical harmonics functions $\mathbf Y_{lm}$ acting on the vector input $\vec r_{ij}$ and $\vec m_i$ are used for equivariant features with arbitrary $l$. The ENN iteratively updates the equivariant features for vertices $\mathbf v_i$ and edges $\mathbf e_{ij}$ by updating them using features in their neighborhood. Contributions from different atoms are aggregated by summation such that the updated features are invariant with respect to atomic permutations. Handling invariance under translation of atoms is obvious, since we only deal with the relative position between atoms and atomic magnetic moments, both of which are invariant with respect to translation.

An overview of the neural-network architecture is presented in Fig. \ref{fig2}. Equivariant features are updated using building blocks in DeepH-E3~\cite{Gong2022}. Importantly, interaction between different angular momentum quantum number $l$ and different atom $i$ is implemented using a tensor product layer~\cite{Thomas2018}: $\mathbf z^{l_3}_{m_3}=\sum_{m_1, m_2}C_{m_1m_2m_3}^{l_1l_2l_3}\mathbf x^{l_1}_{m_1}\mathbf y^{l_2}_{m_2}$. xDeepH embeds atomic structure as initial vertex and edge features, followed by successive vertex layers and edge layers to update corresponding features. Distant information of atomic structure $\{\mathcal R\}$ is aggregated into features upon successive stacking of layers. The influence of local magnetic moment on DFT Hamiltonian is more localized. Regarding this locality, we introduce magnetic configuration $\{\mathcal M\}$ to xDeepH in a strictly localized manner, as illustrated in the magnetic moment layer in Fig. \ref{fig2}. We find that introducing $\{\mathcal M\}$ in this strictly localized manner makes training more efficient and accurate compared to treating $\{\mathcal R\}$ and $\{\mathcal M\}$ on an equal footing (see method details and ablation studies in SM~\cite{supp}). Finally, equivariant features on edges $\mathbf e_{ij}$ are used to construct the Hamiltonian matrix block $H_{ij}$. The xDeepH model is trained using DFT data of small structures by minimizing the loss function defined as the mean-squared errors of the DFT Hamiltonian matrix elements.

The capability of xDeepH is tested by example studies of CrI$_3$ and NiBr$_2$. Our results demonstrate that xDeepH can well reproduce DFT results. Remarkably, once trained by DFT data on small structures with random magnetic orientation, xDeepH can make accurate predictions on large-scale structures with complex magnetic configurations. To generate the dataset and benchmark results, we calculate DFT Hamiltonians for given magnetic configurations using constraint DFT as implemented in the OpenMX package~\cite{Ozaki2003, Ozaki2004, Kurz2004}.

CrI$_3$ nanotubes are interesting for the study of curved magnetism~\cite{Edstrom2022}, but limited by their large-size structures and diverse magnetic configurations. It is difficult to study electronic structure properties directly by DFT calculation for such kind of systems. xDeepH is aimed to solve the problem. Modeling $H_{\text{DFT}}\left(\mathcal R, \mathcal M\right)$ with deep neural networks is a difficult task when both $\mathcal R$ and $\mathcal M$ vary simultaneously, but can be handled by xDeepH as demonstrated in the example study below. 100 different atomic structures of 2$\times$2 supercells are prepared by introducing random atomic displacements (up to  0.1 \AA) to each atom. For each atomic structure, we generate 10 random magnetic configurations by arbitrarily arranging the orientation of the constraint magnetic moment for each magnetic atom Cr. In total, 1000 supercell configurations with random $\{\mathcal R\}$ and $\{\mathcal M\}$ are included in the datasets for CrI$_3$ and are divided into training, validation and test sets with a ratio of $6:2:2$. The mean absolute error (MAE) of the DFT Hamiltonian matrix for configurations in test set is as low as sub-meV (see SM~\cite{supp}). Due to the accurate prediction of the DFT Hamiltonian, band structures obtained from xDeepH agree well with DFT results (see SM~\cite{supp}).

The generalization ability of xDeepH is demonstrated by efficient and accurate predicting of new structures unseen in the training set. (16, 16) armchair CrI$_3$ nanotubes with the non-collinear magnetization normal to the surface or collinear ferromagnetism are selected to test xDeepH [Fig. \ref{fig3}(a)]. The predicted band structures can match the DFT benchmark calculations well [Fig. \ref{fig3}(b)].

We further test the performance of xDeepH to study long-range magnetic order in monolayer NiBr$_2$. It is challenging for DFT to study the long-wavelength spin-spiral magnets~\cite{Tokunaga2011}. 500 3$\times$3 supercells of NiBr$_2$ with equilibrium atomic positions and random orientations of magnetic moments of Ni atoms [Fig. \ref{fig3}(c)]. We split the dataset into training, validation and test sets with a ratio same as that of CrI$_3$. Satisfactory results can be obtained on the test set (see SM~\cite{supp}). The trained xDeepH model can be used to make predictions on NiBr$_2$ with spiral magnetic structure and reproduce DFT band structures [Fig. \ref{fig3}(d)]. These experiments demonstrate good generalization ability of xDeepH, which is useful for investigating  magnetic superstructures.

\begin{figure}
\includegraphics[width=1\linewidth]{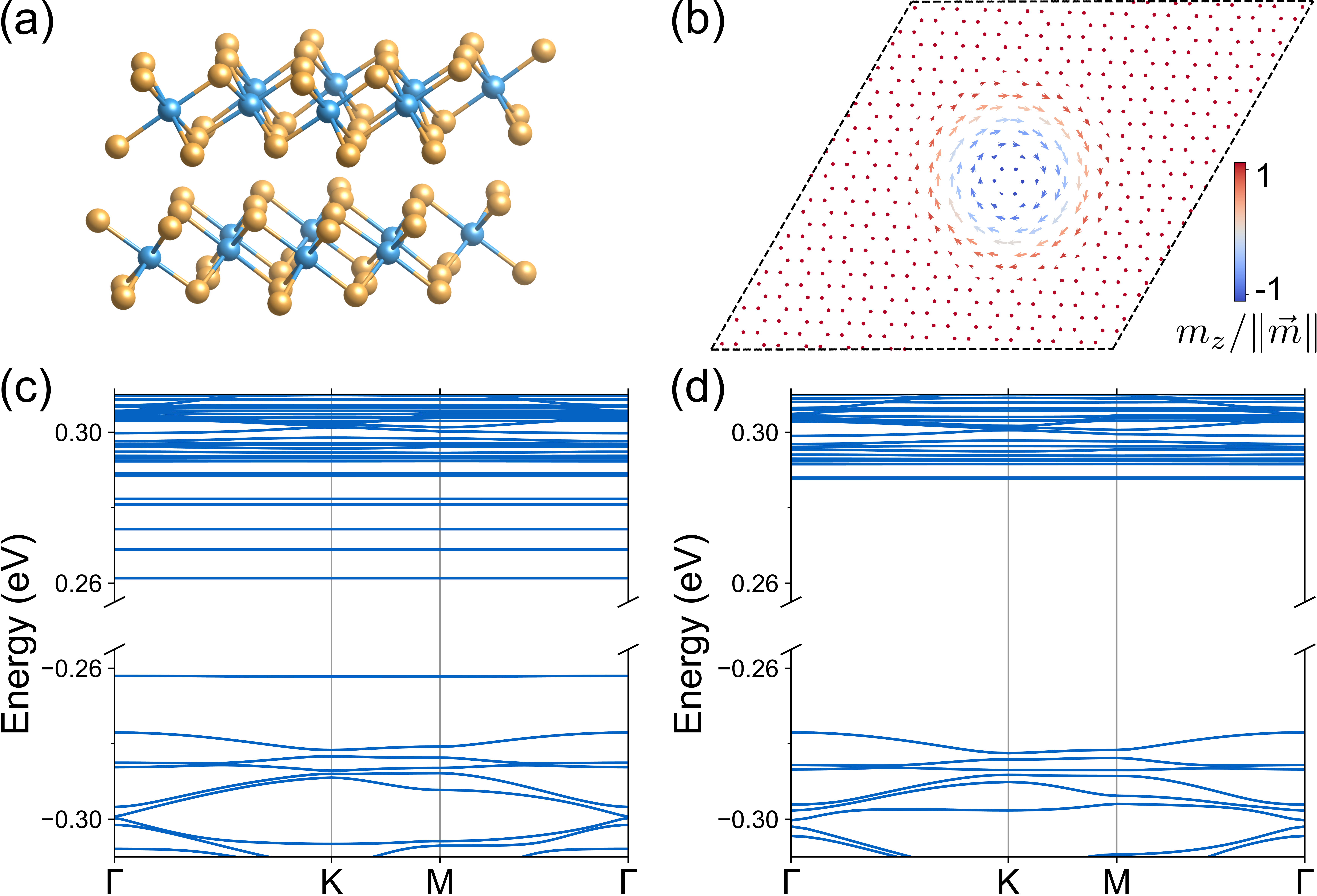}
\caption{Application of xDeepH to study Moir\'{e} magnets without or with magnetic skyrmion. (a) Schematic structure of bilayer CrI$_3$. (b) Magnetic skyrmion in the Moir\'{e}-twisted bilayer CrI$_3$ (twist angle 63.48$^\circ$ and 4,336 atoms per supercell)~\cite{Zheng2022}. Magnetic moments of the top CrI$_3$ layer are labelled by colored arrows, whose out-of-plane components are shown by the color. The underlying CrI$_3$ layer is in the ferromagnetic configuration with up magnetic moments. (c,d) Band structures of the Moir\'{e}-twisted bilayer CrI$_3$ (c) in the ferromagnetic configuration and (d) in the magnetic skyrmion configuration.}
\label{fig4}
\end{figure}

Finally, we apply the xDeepH method to study the Moir\'{e}-twisted bilayer CrI$_3$ [Fig. \ref{fig4}(a)]. The neural network models are trained by DFT datasets obtained for sample structures with a fixed twist angle $\theta = 60 ^\circ$, and then applied to study Moir\'{e}-twisted structures of varying $\theta$. Band structures and density of states obtained from xDeepH match well with the DFT results for a different twist angle $\theta = 81.79 ^\circ$ (see SM~\cite{supp}), verifying reliability of the approach. Next, we consider a twist angle of $\theta = 63.48 ^\circ$, which has magnetic skyrmion as shown in [Fig. \ref{fig4}(b)]~\cite{Zheng2022}. In the ferromagnetic configuration, an extremely flat valence band emerges in the system [Fig. \ref{fig4}(c)], which is originated from Moir\'{e} twist. Interestingly, the flat band is removed away from the valence band edge by magnetic skyrmion [Fig. \ref{fig4}(d)], indicating strong coupling between flat band and magnetic skyrmion. The intriguing physics relevant to flat band and magnetic skyrmion will be further explored in our ensuing work.

In summary, we have proposed a general framework to represent the DFT Hamiltonian of magnetic materials by deep neural networks, which builds a mapping from atomic structures and magnetic configurations to physical properties. All known physical symmetries of the DFT Hamiltonian can be handled by xDeepH, which significantly reduces training complexity and the amount of training data required. High accuracy and satisfactory generalization ability of the method are demonstrated by example studies of CrI$_3$ and NiBr$_2$ systems. This approach opens opportunities to study novel magnetism and electron-magnon coupling in large-scale material systems. Furthermore, one may combine xDeepH with effective spin models or machine-learning potentials of magnetic systems~\cite{Novikov2022, Yu2022} to study the influence of spin dynamics on electronic properties.

\begin{acknowledgements}
We thank F. Zheng for providing the magnetic skyrmion structures. This work was supported by the Basic Science Center Project of NSFC (grant no. 51788104), the National Science Fund for Distinguished Young Scholars (grant no. 12025405), the National Natural Science Foundation of China (grant no. 11874035), the Ministry of Science and Technology of China (grant nos. 2018YFA0307100 and 2018YFA0305603), the Beijing Advanced Innovation Center for Future Chip (ICFC), and the Beijing Advanced Innovation Center for Materials Genome Engineering.

H.L. and Z.T. contributed equally to this work.
\end{acknowledgements}

\end{document}